\documentclass{PoS}

\newcommand{\mnras} {MNRAS}

\newcommand{\apj}{ApJ}

\newcommand{\apjs}{ApJS}
\newcommand{\prd}{PhRvD}
\newcommand{\aap}{A\&A}

\title{Multi-wavelength constraints on cosmic-ray leptons in the Galaxy}

\ShortTitle{Cosmic rays from multi-wavelength observations}

\author{{\speaker{E.~Orlando}~$^a$, A.~W.~Strong$^b$, I. V.~Moskalenko$^a$, C. Dickinson$^c$, S.~Digel$^a$, T.~R. Jaffe$^d$ G.~J\'ohannesson$^e$, J.~P. Leahy$^c$,   T.~A.~Porter$^a$, M. Vidal$^c$ }\\
\newline
\llap{$^a$}Hansen Experimental Physics Laboratory and Kavli Institute for Particle Astrophysics and Cosmology,
        Stanford University, SLAC National Accelerator Laboratory, Stanford, CA 94305, U.S.A. E-mail: \email{eorlando@stanford.edu}\\
\llap{$^b$}Max-Planck-Institut f\"ur extraterrestrische Physik, Postfach 1312, D-85741 Garching, Germany\\
\llap{$^c$}Jodrell Bank Centre for Astrophysics, Alan Turing Building, School of Physics and Astronomy, The University of Manchester, Oxford Road, Manchester, M13 9PL, U.K. \\
\llap{$^d$}CNRS; UPS-OMP; IRAP; Toulouse, France\\
\llap{$^e$}Science Institute, University of Iceland, Dunhaga 5, IS-107 Reykjav\'ik, Iceland\\
        }

\abstract{Cosmic rays (CRs) interact with the gas, the radiation field and the magnetic field in the Milky Way, producing diffuse emission from radio to gamma rays. Observations of this diffuse emission and comparison with detailed predictions are powerful tools to unveil the CR properties and to study CR propagation. We present various GALPROP CR propagation scenarios based on current CR measurements. The predicted synchrotron emission is compared to radio surveys, and synchrotron temperature maps from {\it WMAP} and {\it Planck}, while the predicted interstellar gamma-ray emission is compared to {\it Fermi}-LAT observations. We show how multi-wavelength observations of the Galactic diffuse emission can be used to help constrain the CR lepton spectrum and propagation. Finally we discuss how radio and microwave data could be used in understanding the diffuse Galactic gamma-ray emission observed with {\it Fermi}-LAT, especially at low energies.}

\FullConference{The 34th International Cosmic Ray Conference,\\
		30 July- 6 August, 2015\\
		The Hague, The Netherlands}

\begin{document}

\section{Cosmic ray leptons and diffuse emission} 
Below a few GeV the local interstellar lepton spectrum (LIS) cannot be directly measured, because CRs are affected by solar modulation. At higher energies, CR measurements might not be representative for the average spectrum in the Galaxy \cite{strong2004}. During propagation in the Galaxy CR leptons are interacting with the interstellar medium (ISM), and are scrambled by the magnetic field (B-field). Their spectrum is steepened by energy losses and energy-dependent diffusion, and it can also be affected by re-acceleration and production of secondary CRs \cite{strong2007}. Hence, tracing CR leptons back to their origin and  throughout the Galaxy is difficult. However, a way to study CR leptons in the Galaxy is to observe the diffuse emission produced by their interaction with B-fields via synchrotron emission (in radio and microwave) and with the gas and the interstellar radiation field (ISRF) via bremsstrahlung and inverse Compton (IC) emission (in gamma rays). These observations, compared with sophisticated propagation models, allow us to gain information on the CR propagation in the Galaxy and and their interstellar spectrum. In particular CR leptons from $\sim$500\,MeV to tens of GeV produce synchrotron emission from tens of MHz to hundreds of GHz for a typical B-field of a few $\mu$G \cite{SOJ2011}. The observed synchrotron spectral index depends on the spectral indices of CR leptons along the line of sight. Hence this can be used in conjunction with direct measurements to construct the full interstellar lepton spectrum from hundreds of MeV to TeV. Combining observations of the diffuse emission from radio through
gamma rays, with the aid of propagation models, then allows us to derive the quantities related to the CRs, such as spectrum, density, distribution in the Galaxy, and propagation parameters. This approach has certain advantages  over studying these data separately by providing independent information on
the spectra of CR leptons, and hence reducing degeneracy.

After presenting a summary of our previous studies on CR leptons and diffuse emissions, we account here for recent PAMELA CR lepton measurements \cite{Pamela}. The derived synchrotron emission is calculated and its spectrum is compared with radio surveys and recent observations from the 9-year {\it WMAP} \cite{bennett} and four-year {\it Planck} synchrotron temperature maps \cite{Ingunn}. The gamma-ray diffuse emission is calculated and compared to the {\it Fermi} Large Area Telescope (LAT) data. 

\section{Modeling CRs with GALPROP}
This study uses a numerical model of CR propagation and interactions in the Galaxy, GALPROP\footnote{http://galprop.stanford.edu/}. Descriptions of the code can be found in \cite{strong2007, moskalenko98, vladimirov} and references therein. 
It enables simultaneous predictions of observations of CRs, gamma rays and synchrotron radiation.
GALPROP models were used to analyze and interpret the diffuse gamma rays detected by EGRET, COMPTEL (e.g. \cite{strong2004}), {\it INTEGRAL} (e.g. \cite{porter2008}), LAT  (e.g. \cite{diffuse1, diffuse2,  JM, Tibaldo, Orlando, Dermer}), and more recently radio and microwave observations (e.g. \cite{SOJ2011, O&S2013}). 
GALPROP calculates the diffuse emission from pion decay, bremsstrahlung and IC, for a user-defined
CR source distribution and propagation parameters.
Sample gas maps and ISRF are given in \cite{diffuse2}. 
Calculation  of interstellar synchrotron emission has been improved \cite{SOJ2011} and extended \cite{O&S2013} to include 3D B-field models, synchrotron polarization and a basic model for free-free emission and absorption. Also 3D gas and source distribution models were recently implemented \cite{Gulli} and J{\'o}hannesson et al. (these Proceedings). For an updated overview see \cite{igor2015icrc} and Moskalenko et al. (these Proceedings).

\begin{figure}
\centering
\includegraphics[width=20pc, angle=0]{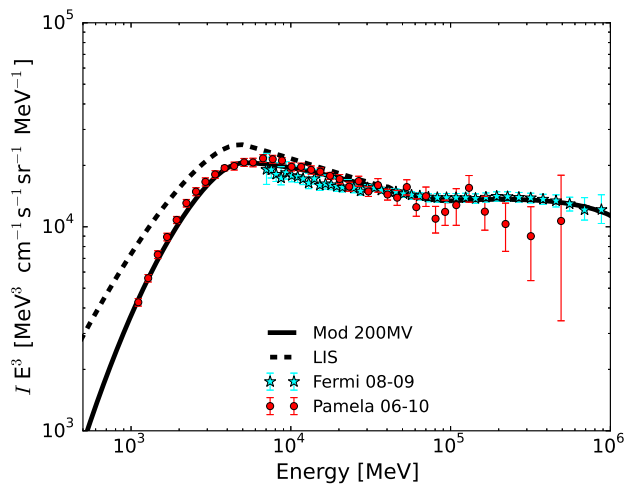}
\caption{Modeled LIS for a plain diffusive propagation model and modulated spectrum compared with PAMELA \cite{Pamela} and LAT \cite{fermi_ele} measurements, taken mostly during solar minimum. }
\label{fig1}
\end{figure}

\begin{figure}
\centering
\includegraphics[width=20pc, angle=0]{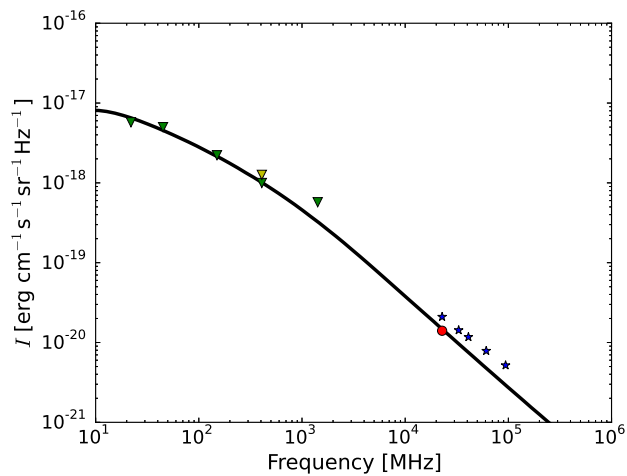}
\caption{Synchrotron spectrum ({\it black line}) from high latitudes ($10^{\circ}<|b|<45^{\circ}$, $0^{\circ}<l<360^{\circ}$) for the plain diffusive propagation model calculated with the CR leptons as in Fig.\,\protect\ref{fig1}.  Data are:  9-year {\it WMAP} synchrotron temperature maps ({\it blue stars}) \cite{bennett}; {\it Planck} synchrotron maps scaled to 23\,GHz \cite{Paddy} ({\it red point}); radio surveys described in the text and in \cite{SOJ2011} ({\it green triangles}); 408\,MHz from \cite{haslam81} with 3.7\,K offset ({\it yellow triangle}) and reprocessed 408\,MHz from \cite{408new} with 8.9\,K offset ({\it green triangle}).}
\label{fig2}
\end{figure}

\section{Cosmic ray leptons: summary of our previous studies}
In our recent papers \cite{SOJ2011, diffuse2, O&S2013, jaffe2013}, AMS-01 \cite{AMS} measurements at low energies and LAT \cite{fermi_ele} data at higher energies
were used to derive the spectral index of the CR lepton spectrum. Unfortunately, CR measurements are not sufficient to constrain the leptons at low energies, since there is a strong correlation in the models between the spectral index and the modulation potential. Hence, to break this degeneracy we derived the local spectral index at low energies from synchrotron observations \cite{SOJ2011}, using a collection of radio surveys and spectral index data over a wide range of frequencies and the synchrotron temperature maps from \cite{MD}. This allowed us to constrain the LIS at low-energies and propagation models
that best fit the observations while avoiding the effects of solar modulation.
We concluded that the interstellar spectrum turns over rather sharply below a few GeV. This is independent from propagation effects, and it reflects the electron injection spectrum from the sources.
The low-energy falloff in the directly measured electrons, normally attributed mainly to modulation, is partially due to the interstellar spectrum. This suggests less solar modulation than usually assumed. We found that it is challenging to describe the observed synchrotron spectrum using present diffusive re-acceleration
models with a standard Alfven velocity (that works fine for other CR species),
since the modeled emission from primary and secondary leptons exceeds the measured synchrotron below hundreds of MHz. 
Spatial effects, independently in gamma rays with the LAT \cite{diffuse2} and in radio and microwaves \cite{O&S2013}, were extensively investigated, using different CR distributions, propagation halo sizes, B-field and ISM models. We report here on the spectral properties only, updating our previous analyses by using recent PAMELA CR measurements, the reprocessed 408\,MHz map \cite{408new}, the new 9-year {\it WMAP} maps, the recently released {\it Planck} maps, and gamma-ray observations with the LAT. 

\section{Description of our approach}
We use GALROP for modeling CR propagation requiring that the modulated spectrum reproduces
CR measurements. Our method is to extract information on the spectral indexes of the injected particles
and propagation parameters by simultaneously studying the radio, microwave and gamma-ray emissions produced by
CR leptons.
This allows us to bypass part of the degeneracy of the parameters, such as solar modulation, primary electron spectral index and contribution of secondaries. Specifically, first we derive constraints on the LIS from direct CR measurements. Then we calculate the diffuse synchrotron emission in the radio and microwave band and compare with synchrotron
observations with the aim of tuning the lepton spectrum at low energy to match the observations
when possible. Then we use this information to generate the leptonic diffuse gamma-ray emission.
In turn, comparison of the predictions of the diffuse gamma-ray emission with LAT observations provide us
additional information on CR leptons and propagation. Therefore, gamma-ray and synchrotron
emission can probe the interstellar spectrum and propagation effects free from the effects of modulation.

\section{Observations}
We use the surveys at 22, 45, 150, 408, and 1420\,MHz and their zero levels as in \cite{SOJ2011}. 
Frequencies from 20\,MHz to a few GHz account mostly for the synchrotron emission. The emission there is produced primarily by electrons with energy below a few GeV. 
The reprocessed 408\,MHz map by \cite{408new} with better source subtraction is also used. Its offset is taken from \cite{Ingunn408} as 8.9\,K.
Thanks to the latest release this year \cite{mapoverview}, we use here the {\it Planck} low frequency separated component maps. 
In addition the seven-year {\it WMAP} temperature maps used in \cite{O&S2013} are updated here with the 9-year {\it WMAP} synchrotron maps from 23 to 94\,GHz obtained with the Maximum Entropy Method (MEM).
We use LAT observations as published in [11], from intermediate latitudes that trace relatively local CRs. The used LAT data include 21 months of observations and Pass 6 DataClean
event selection.

\section{Propagation models}
We follow \cite{SOJ2011} and start from the same plain diffusive model that fits synchrotron observations. We tune the propagation parameters from B/C, protons, and helium to PAMELA measurements. Then we tune the electrons to direct LAT measurements above 100 GeV where the solar modulation is negligible. The modulated low energy spectrum is then tuned to reproduce simultaneously PAMELA CR measurements and synchrotron observations, and finally we compare the models to gamma-rays to make sure the modeled emission matches data. The local CR lepton spectrum for the new propagation model is shown in Fig.\,\ref{fig1}. Note that PAMELA measures the electron spectrum, and its statistical errors increase with energy. At higher energies we reproduce the total lepton spectrum
as measured by the LAT. The latter combines electrons plus positrons so that the increasing positron
fraction at high energies is included in our modeling. 

\section{Results: synchrotron and gamma-ray  spectra}
\begin{figure}
\centering
\includegraphics[width=20pc, angle=0]{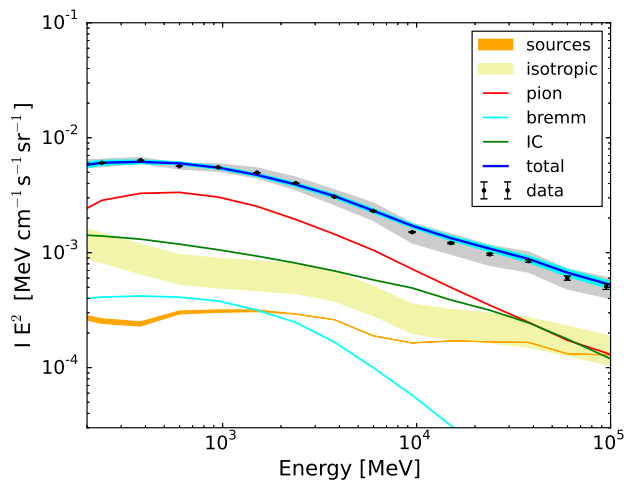}
\caption{Calculated gamma-ray components and spectrum compared with the LAT data from \cite{diffuse2} for intermediate latitudes ($10^{\circ}<|b|<20^{\circ}$, all longitudes), using the plain diffusive propagation model and the LIS as in Fig.\,\protect\ref{fig1}. Data include statistical ({\it grey area}) and systematic errors ({\it black bars}). Here components are not fitted to gamma-ray data, so uncertainties in the gas and ISRF are not accounted for. Spectra for sources and isotropic are taken as in \cite{diffuse2}, for the most extreme cases reported there.  30$\%$ uncertainty is added to the isotropic spectrum, following the study in \cite{EGB} based on various foreground models. The ISRF has been scaled to better reproduce gamma rays as found in \cite{diffuse2}.}
\label{fig3}
\end{figure}

\begin{figure}
\centering
\includegraphics[width=16pc, angle=0]{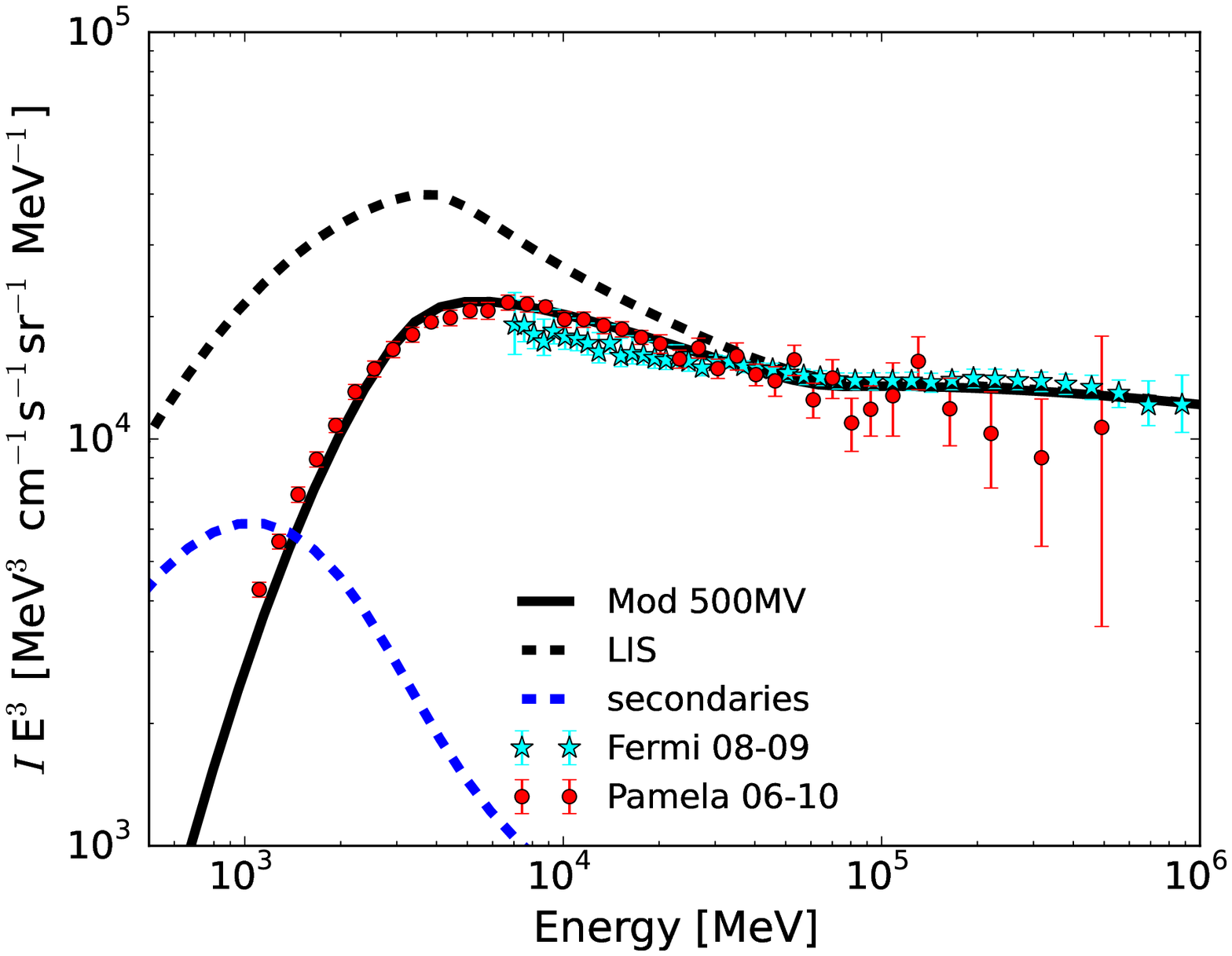}
\includegraphics[width=16pc, angle=0]{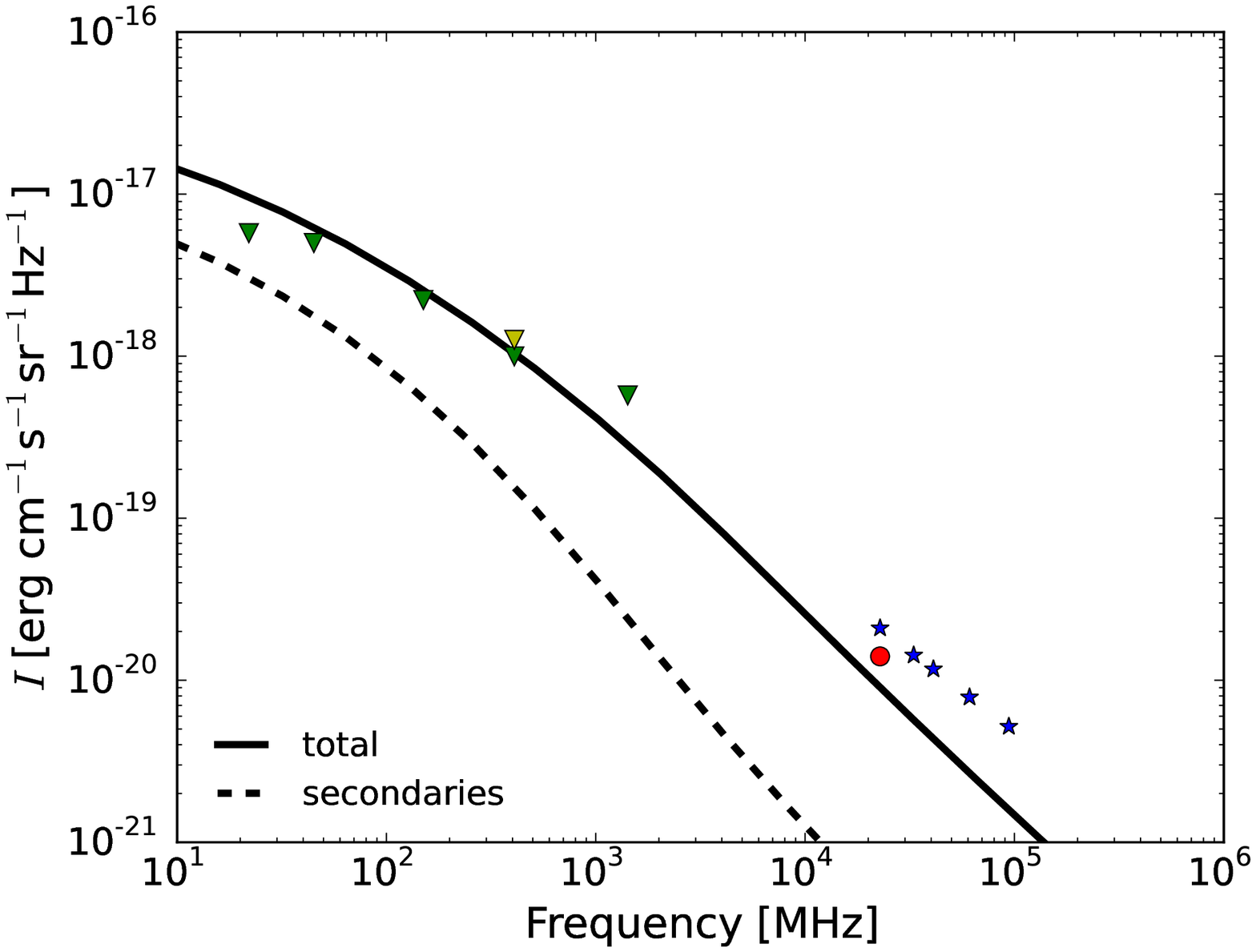}
\caption{{\it Left}: modeled LIS and modulated spectrum for a diffusive re-acceleration model from \cite{Vlad}. Data are as in Fig.\,\protect\ref{fig1}. The blue line is the model component for secondary leptons. {\it Right}: Calculated synchrotron spectrum for high latitudes compared with data as in Fig.\,\protect\ref{fig2}. Dashed line is the contribution from secondaries.}
\label{fig4}
\end{figure}

Figures \ref{fig2} and \ref{fig3} show results for the plain diffusion models with the LIS as shown in Fig.\,\ref{fig1}. The intensity of the B-field has been fit to the 408\,MHz map from \cite{408new}. The modeled synchrotron spectrum for the high-latitude regions reproduces the observations quite well. Below a few GHz data from surveys are well reproduced by the model. At a few GHz contamination from free-free in this region, not accounted here, is below 25$\%$ \cite{O&S2013, Clive2003}. 
In microwave we see a good spectral agreement with {\it WMAP} maps integrated over the same region of the model, and a good fit to the {\it Planck} component-separated synchrotron map. This comparison shows that the LIS used here is a good representation of the spectrum that produces the synchrotron emission. This suggests that low-energy spectral index, solar modulation and the contribution of secondaries are well constrained, as previously found by \cite{SOJ2011}. The induced gamma-ray emission is shown in Fig.\,\ref{fig3} and compared to the LAT data for the intermediate latitudes taken from \cite{diffuse2}. The model is within the LAT systematic uncertainties even without the tuning to the data that would account for uncertainties in the ISM. Hence, in a first approximation plain diffusive propagation models can reproduce gamma rays as well as the diffusive re-acceleration models usually assumed. A detailed fitting procedure will be described in a separate work. For comparison with published re-acceleration models, Fig.\,\ref{fig4} (left) shows the electron spectrum used in \cite{Vlad} compared with PAMELA measurements. Similar re-acceleration models have been used in \cite{diffuse2} to fit gamma-ray data. However, these models were not tuned to fit synchrotron spectral observations. In fact the same figure also shows the calculated synchrotron spectrum compared with the same observations reported in Fig.\,\ref{fig2}, and with the B-field intensity fit to the 408\,MHz map, as above. An excess of the model at the lowest frequencies is visible, mostly due to secondary leptons produced by re-acceleration processes. This excess is now reduced with respect to \cite{SOJ2011} due to a combination of the reprocessed 408\,MHz map and the larger offset used here. The spectrum does not reproduce the microwave observations as well as the one in Fig.\,\ref{fig2}, since it under-estimates the synchrotron intensity from {\it Planck}. 

\section{Discussion}
In this work we show the feasibility and importance of using multi-wavelength observations, especially at radio wavelengths, together with CR measurements, to constrain the interstellar spectrum at low-energies, propagation models, and solar modulation effects. We have presented preliminary results and now discuss here the limitations and potential of this approach.

The exact derivation of the synchrotron maps as obtained by {\it Planck} and {\it WMAP} have limitations, due to the various assumptions required and degeneracies with separating multiple astrophysical components including synchrotron, free-free, thermal dust and anomalous microwave emissions (AME) (\cite{Ingunn,Paddy}). Disentangling these components in the  {\it WMAP} and {\it Planck} bands requires various assumptions or a priori information such as data-driven templates using ancillary data (where each different emission mechanism is at its maximum) or by approximating their spectral properties (see \cite{bennett, Ingunn, Paddy, Vidal}). The method used to derive the {\it WMAP} MEM temperature maps shown here utilized spatial templates of different emission components from external data as priors, and  the fit was performed to each pixel independently. The spectra for the different components were fixed, while the synchrotron spectrum was obtained from polarization observations. The resulting synchrotron map is plausibly contaminated by other components, such as AME and free-free emission. This could explain the difference with the {\it Planck} data \cite{Paddy}. In fact the intensity of the {\it Planck} synchrotron map is lower than previously found with {\it WMAP}, while the AME and free-free intensities are higher. This might suggest that the {\it Planck} synchrotron intensity could be under-estimated, while the {\it WMAP} intensity is overestimated. 

The high-quality {\it Planck} component maps recently released were combined with complementary ancillary data: the 9-year {\it WMAP} temperature maps and the 408\,MHz map, providing an excellent fit to the data. The synchrotron spectral index from the best model by \cite{O&S2013} was used as the baseline model to allow a better separation of the synchrotron component. Nevertheless, there is likely to be a degeneracies between the various low-frequency components, especially between AME and synchrotron. Hence, these component-separated maps face uncertainties related to our limited knowledge of their spectra and spatial distributions (plus uncertainties in the offset at the different frequencies), as discussed in detail in \cite{Paddy}. Note that also the zero levels of the surveys are not clearly determined, that could limit our model constraints. \cite{Ingunn408} estimated a monopole of 8.9$\pm$1.3\,K in the 408\,MHz map \cite{haslam81}, which includes any isotropic component (CMB, Galactic and extragalactic), which we use for the fit here. However, the determination of the offsets strongly depends on the foreground models and data used. In \cite{SOJ2011} we adopted a 3.6\,K offset, which increased the excess at lower frequencies for the diffusive re-acceleration models. Further model-dependent studies and forthcoming data at a few GHz up to $\approx 15$\,GHz (e.g. CBASS \cite{CBASS}) will help in separating the components and may provide more strict constraints to the lepton spectrum. 

Regarding gamma rays, the sky above 100\,MeV is dominated by emission produced by CRs interacting with the gas and ISRF via pion-decay, IC, and bremsstrahlung. Disentangling the different components at the LAT energies is challenging and is usually done in a model-dependent approach. Uncertainties in the ISM is the major limitation to our modeling and hence in our knowledge of CRs, e.g. as found in \cite{diffuse2}. The situation below 100\,MeV is still unexplored. Extrapolations of present models to such low energies predict IC and bremsstrahlung to be the major mechanisms of CR-induced emission, which are of leptonic origin. The fact that energies $<100$\,MeV were not deeply investigated after the COMPTEL era, makes it more exiting now with the advent of the LAT pass 8 data and its extension to lower energies \cite{Orlando, Pass8}. However disentangling the different components and characterizing the sources below 100\,MeV is even more challenging due to the relatively large point spread function of the instrument. This highlights the importance of multi-wavelength observations including the Square Kilometre Array telescope (e.g. \cite{Clive}) and C-BASS \cite{CBASS} that will provide supplementary information at radio frequencies.

\footnotesize{{\bf Acknowledgment:}{
E.O. acknowledges support via NASA Grant No. NNX13AH72G. GALPROP development is supported through NASA Grants No. NNX10AE78G, 
and NNX13AC47G. C.D. is supported by an ERC Starting Consolidator Grant (no.~307209) and an STFC Consolidated Grant (ST/L000768/1).}}

\end{document}